\documentclass[aps,pra,reprint,showpacs,floatfix,amsmath,amssymb,10pt]{revtex4-1} 

\usepackage[pdftex]{graphicx}       

\usepackage{epstopdf}

\usepackage{amsmath}				

\newcommand{\dd}{\mathrm{d}}
\newcommand{\pd}{\partial}
\newcommand{\hb}{\hbar}

\begin{document}


\title{Universality in rotating strongly interacting gases}

\author{B. C. Mulkerin} 
\author{C. J. Bradly} 
\author{H. M. Quiney}
\author{A. M. Martin}
\affiliation{School of Physics, University of Melbourne, Victoria 3010, Australia}

\date{\today}
\begin{abstract}
We analytically determine the properties of two interacting particles in a harmonic trap subject to a rotation or a uniform synthetic magnetic field, where the spherical symmetry of the relative Hamiltonian is preserved. Thermodynamic quantities such as the entropy and energy are calculated via the second order quantum cluster expansion. We find that in the strongly interacting regime the energy is universal, however the entropy changes as a function of the rotation or synthetic magnetic field strength.
\end{abstract}

\maketitle

Over the last few years ultracold degenerate gases have attracted much interest due to their controllability and stability. Advances in tight confining harmonic traps and the use of magnetic fields and Feshbach resonances in controlling atomic collisions have made it possible to explore the BCS-BEC crossover \cite{Chin2010,Shin:2008,zwierlein-2006-442}. Difficulties with developing a many-body theory for these systems in the strongly interacting regime using mean-field approximations have motivated the study of few-body problems as a means to gain insight into the many-body problem. Few-body systems with contact interactions are exactly solvable or numericaly tractable \cite{Busch:1998,Werner2006,Kestner2007,Daily2010}, particularly in the strongly interacting regime and have been experimentally studied in their own right \cite{Staferle2006}. The virial expansion of few-body physics can be used to calculate the thermodynamics of many-body systems \cite{Liu2009,Liu2010,Daily2012} and has been verified experimentally \cite{Nascimbene2010}. 

In this work we address the problem of unitary gases subject to a rotation or synthetic magnetic field by solving the two-body problem and finding the virial expansion to second order. This enables us to show that entropy in the presence of a rotation or synthetic magnetic field is not universal, in contrast to the universal character of the total energy.



A system subject to a rotation and one subject to a synthetic magnetic field have several similarities. In both systems angular momentum states and time-reversal symmetry are broken. Furthermore, both problems can be described by gauge-dependent Hamiltonians, making it convenient to consider the systems together and to draw comparisons between the two. In ultracold trapped gases the dominant contribution to the low energy behavior is from the two-particle s-wave interactions.

 To begin the analysis the rotating system is considered first. Specifically, the motion of two particles of mass $m$ in a harmonic trap potential $V_\text{trap}(\mathbf{r})$ subject to a rotation $\mathbf{\Omega}$ and a contact interaction potential $V_\text{int}(\mathbf{r_1-r_2})$ are described by the Hamiltonian 
\begin{alignat}{1} \label{eq:2ParticleHamiltonian}
H= & \frac{\mathbf{p}_{1}^2}{2m}+  
\frac{\mathbf{p}_{2}^2}{2m} + V_\text{trap}(\mathbf{r_1}) + V_\text{trap}(\mathbf{r_2}) + \nonumber \\ &
\mathbf{\Omega\cdot r_1 \times p_1}+\mathbf{\Omega\cdot r_2 \times p_2} + V_\text{int}(\mathbf{r_1-r_2}),
\end{alignat} 
where $\mathbf{r_i}$ and $\mathbf{p_i}$ are the positions and momenta of each particle. Equation \eqref{eq:2ParticleHamiltonian} can be decoupled in center of mass and relative coordinates, yielding
\begin{alignat}{1}
H_\text{cm}&=\frac{\mathbf{P}^2}{4m}+V_\text{trap}(\mathbf{R})-\mathbf{\Omega \cdot R \times P} \label{eq:COMHamiltonian} \\  
H_\text{rel}&=\frac{\mathbf{p}^2}{m}+V_\text{trap}(\mathbf{r})-\mathbf{\Omega \cdot r \times p}+V_\text{int}(\mathbf{r}), 
\label{eq:RelativeHamiltonian}
\end{alignat}
where $\mathbf{R}=(\mathbf{r_1+r_2})/\sqrt{2}$ and $\mathbf{r}=(\mathbf{r_1-r_2})/\sqrt{2}$ are the center of mass and relative coordinates. We consider the case where the rotation is about the $z$-axis with frequency $\Omega_z$ so that $\mathbf{\Omega}=(0,0,\Omega_z)$. The harmonic trapping potential is chosen to be axially symmetric 
with transverse and axial frequencies $\omega_\perp$ and $\omega_z$, respectively. Using the axial trap length $d=\sqrt{\hbar/(m\omega_z)}$ and energy $\hbar\omega_z$, Eq.~\eqref{eq:COMHamiltonian} can be written in the dimensionless form 
\begin{alignat}{1}\label{eq:DimensionlessHamiltonian}
H_\text{cm}=\frac{1}{2}\nabla^2_\mathbf{R} + \frac{1}{2}\left(\eta^2 \rho^2+z^2\right) - i\xi\left(x\frac{\partial}{\partial y}-y\frac{\partial}{\partial x}\right),
\end{alignat}
where $\eta=\omega_\perp/\omega_z$ and $\xi=-\Omega_z/\omega_z$ parameterize the aspect ratio of the trap and the applied rotation.
Equation \eqref{eq:DimensionlessHamiltonian} is the Hamiltonian for a shifted anisotropic harmonic oscillator with eigenstates
\begin{alignat}{1}
& \psi_{nmk}(\rho,\phi,z)=R_{nm}(\rho,\phi) Z_k(z) \label{eq:Basis} \\ & 
R_{nm}(\rho,\phi) =  \sqrt{\frac{\eta^{|m|+1} n!}{(n+m)!\pi}} \rho^{|m|}  e^{-\eta\rho^2/2} L^{|m|}_n(\eta\rho^2) e^{im\phi} \\ & 
Z_k(z)=\frac{e^{-z^2/2}}{\pi^{1/4}\sqrt{2^k k!}} H_k(z),
\end{alignat}
and eigenenergies, in units of $\hbar \omega_z$,
\begin{equation} \label{eq:SingleParticleEnergies}
E_{nmk}=(2n+|m|+1)\eta + m\xi + (k+1/2),
\end{equation}
where $L^{|m|}_n(\rho)$ and $H_k(z)$ are, respectively, Laguerre and Hermite polynomials.  The center-of-mass component is therefore solved exactly and the effects of the interparticle interaction are described entirely within the relative component. 

In the presence of a synthetic magnetic field the Hamiltonian for the center-of-mass of two anisotropically trapped particles is
\begin{equation}\label{eq:MagHamiltonian}
H_\text{cm}^\text{mag}=\frac{1}{4m}\left({\bf{P}} - q \bf{A}\right)^2 + V_\text{trap}(\mathbf{R}),
\end{equation}
where $\mathbf{A}$ is the synthetic magnetic vector potential. While the properties of a magnetic system are formally independent of choice of gauge, the problem is not analytically tractable in the Landau gauge, but is tractable in the symmetric gauge. For a uniform synthetic magnetic field, $B$, in the $z$-direction we therefore set $\mathbf{A}=\tfrac{B}{2}(-y,x,0)$. It can then be shown that the eigenstates and eigenenergies for Eq.~\eqref{eq:MagHamiltonian} are given exactly by Eqs.~\eqref{eq:Basis} and \eqref{eq:SingleParticleEnergies} with redfined parameters
\begin{equation}\label{eq:EffectiveFrequencies}
\xi_\text{cm}=\frac{\omega_c}{\omega_z} \quad \text{and} \quad \eta_\text{cm}=\sqrt{\frac{\omega_\perp^2}{\omega_z^2} + \frac{\omega_c^2}{4\omega_z^2}},
\end{equation}
where $\omega_c=|qB/M|$ is the cyclotron frequency with $M=2m$. For the single-particle Hamiltonian, the eigenstates and eigenenergies are expressed in terms of $\xi_1$ and $\eta_1$ using $M=m$ and for the relative Hamiltonian the eigenstates and eigenenergies are expressed in terms of $\xi_\text{rel}$ and $\eta_\text{rel}$ using $M=m/2$. In contrast, for the rotating case the relative, centre-of-mass and single particles are parameterised by the same $\eta$ and $\xi$ as defined in Eq.~\eqref{eq:DimensionlessHamiltonian}. 

Having solved the center of mass Hamiltonian in both the rotating and synthetic magnetic field cases, the relative Hamiltonian needs to be solved exactly. The low-energy regularized $s$-wave contact interaction is \cite{Huang1957}
\begin{alignat}{1}\label{eq:ContactInt}
V_\text{int}(\mathbf{r})=\frac{4\pi\hbar^2a}{m}\delta(\mathbf{r})\frac{\partial}{\partial r}r,
\end{alignat}
where $a$ is the scattering length. In the single-particle basis the relative wavefunction can be written as
\begin{equation} \label{eq:RelativeWaveFxn}
\Psi_\text{rel}({\bf r})=\sum_{n,k=0}^\infty c_{nk} \psi_{n0k}(\rho,\phi,z),
\end{equation}
where the $m \not =0$ states are omitted because they do not contribute, due to the centrifugal barrier in cylindrical coordinates. As such, the projection of angular momentum along the $z$-axis due to the rotation or synthetic magnetic field is contained entirely within the center-of-mass energy. Following \cite{Idziaszek2006} we find that the energy spectrum of the relative motion can be determined from
\begin{equation}\label{eq:Transcendental}
\mathcal{F}\left(-\frac{E-\eta-\tfrac{1}{2}}{2}\right)=-\sqrt{2\pi}\frac{d}{a} ,
\end{equation}
where
\begin{equation}
\mathcal{F}(x)=\int_0^\infty \!\!\!\! \mathrm{d}t \left[ \frac{\eta e^{-x t}}{\sqrt{1-e^{-t}}(1-e^{-\eta t})} - \frac{1}{t^{3/2}} \right].
\end{equation}
This integral is not analytic in general and is formally valid only for $x>0$ but can be extended to all energies by the recurrence relation
\begin{alignat}{1}
\mathcal{F}(x)+\mathcal{F}(x+\eta)=\eta\sqrt{\pi}\frac{\Gamma(x)}{\Gamma(x+1/2)}.
\label{eq:Recurrence}
\end{alignat}
\begin{figure}[bt]
     \includegraphics[width=\columnwidth ]{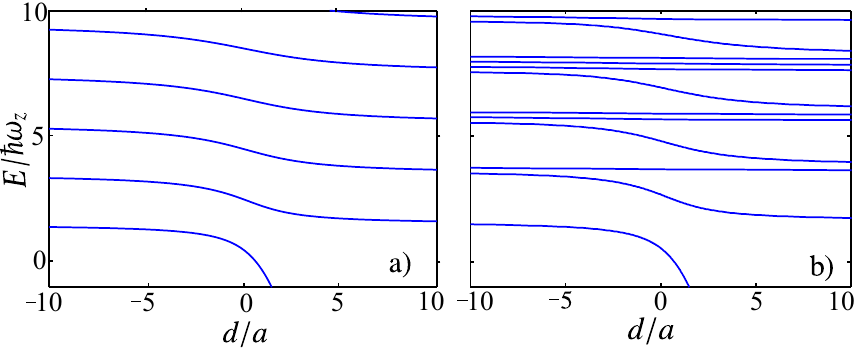}
  \caption{Eigenenergies of the relative Hamiltonian for (a) $\eta=1$ and (b) $\eta=\sqrt{11/9} \approx 1.10$, as a function of the inverse scattering length.}
  \label{fig:RelativeEnergies}
\end{figure}

In Fig.~\ref{fig:RelativeEnergies} we plot the energy eigenspectrum Eq.~\eqref{eq:Transcendental} as a function of inverse scattering length $d/a$ for two values of $\eta$. For the case $\eta=1$, Fig.~\ref{fig:RelativeEnergies}(a) shows states evenly spaced by $2\hbar \omega_z$ with a smooth transition across the Feshbach resonance and a single bound state as the ground-state in the repulsive regime. Unlike in the spherical basis, the $s$-wave interaction allows the orbital angular momentum states to be present at each energy level but they are degenerate. It is possible to take a linear combination of these degenerate wave functions to obtain the usual spherically symmetric, zero orbital angular momentum wave function. If $\eta$ is rational, by the properties of \eqref{eq:SingleParticleEnergies} some of these degeneracies are restored. For the special case when $\eta$, or $1/\eta$, is an integer then all degeneracies are restored \cite{Peng2011}. The most extreme case is when $\eta$ is irrational, for which there can be no degeneracies with higher states. If the spherical symmetry is broken so that $\eta \not =1$ then the degeneracies are lifted, as shown in Fig.~\ref{fig:RelativeEnergies}(b). Angular momentum is no longer conserved and a linear combination of wave functions will not reduce the number of accessible states. As such, one could argue that as soon as the spherical symmetry of the problem is broken the properties of the gas become independent of the relative Hamiltonian and hence the interactions. Of course, this cannot be correct and to overcome this dichotomy one must consider an interaction potential that is regularized in accordance with the symmetries of the system \cite{Wodkiewicz1991}, and including contributions from higher order (non-$s$-wave scattering) processes \cite{Derevianko2003,Pricoupenko2006,Bolda2003,Idziaszek2006a}.

In the following analysis, we assume spherical symmetry of the relative Hamiltonian. In the rotation case, $\eta=1$ means that the trap is isotropic and the relative eigenenergies are unaffected by the rotation. In the synthetic magnetic field case, $\eta_\text{rel}=1$ is equivalent to \smash{$\omega_\perp=[\omega_{z}^2-(qB/m)^2]^{1/2}$}, from Eq.~\eqref{eq:EffectiveFrequencies}. This means that for a given magnetic field there is a particular trap shape that preserves the spherical symmetry of $H_\text{rel}$. 

Few-body physics has importance beyond small scale systems into the thermodynamics of a many-particle gas. We can achieve this through a quantum cluster expansion of the grand thermodynamic potential $\Omega=-k_B T \ln \mathcal{Z}$ in terms of the fugacity $z=\exp(\mu/k_B T)$: 
\begin{equation}
\Omega=-k_B T Q_1 \left(z+b_2 z^2+\dots\right),
\label{eq:GTPOmega}
\end{equation}
where
\begin{equation}
b_2=(Q_2-Q_1^{2}/2)/Q_1
\end{equation}
is the second virial coefficient and the $N$-particle partition function $Q_N=\text{Tr}[\exp(-\mathcal{H}_N/k_B T)]$ is determined from solving the $N$-body problem \cite{HuangBook}. 

In order to calculate the thermodynamics it is more convenient to consider the difference between the interacting and non-interacting systems and define
\begin{equation}\label{eq:Deltab2}
\Delta b_2=(b_2-b_2^{(0)}) = (Q_2 - Q_2^{(0)})/Q_1,
\end{equation}
where the superscript `0' denotes non-interacting quantities. The thermodynamics of a non-interacting two-component Fermi gas in an anisotropic trap with a constant magnetic field or a rotation are determined from the grand potential $\Omega^{(0)}$. Using the energy spectrum \eqref{eq:SingleParticleEnergies} to determine the density of states it can be shown that
\begin{equation}\label{eq:NonIntGrandPotential}
\Omega^{(0)} = -k_B T Q_1^{(0)} \frac{1}{2} \int_0^\infty \!\!\!\! \dd E E^2 \ln(1+z e^{-E}) ,
\end{equation}
where
\begin{equation}\label{eq:NonIntQ1}
Q_1^{(0)} = 2\left(\frac{\hb \omega_z}{k_B T}\right)^3 \left(\frac{1}{2\eta(\eta+\xi)} + \frac{1}{2\eta(\eta-\xi)}\right).
\end{equation}
Equations \eqref{eq:NonIntGrandPotential} and \eqref{eq:NonIntQ1} reduce to the more familiar forms in the limits of no rotation or synthetic magnetic field ($\xi \to 0$) and isotropic trapping ($\eta \to 1$) \cite{Liu2010}. The integral in Eq.~\eqref{eq:NonIntGrandPotential} can be expanded in powers of the fugacity $z$ to obtain the non-interacting virial coefficients \smash{$b^{(0)}_n = (-1)^{n+1}/n^4$}. 

In the thermodynamic limit, $\Delta b_2$ is independent of temperature. In the following calculations it is useful to introduce the small parameter $\tilde{\omega}=\hb\omega_z/k_B T$. Expanding the virial coefficients and cluster partition function in the small parameter $\tilde{\omega}$ allows us to determine their universality.

Since the two-body problem may be separated into relative and center-of-mass coordinates, we can sum over the center-of-mass component $Q_\text{cm}$ and the relative component independently so that Eq.~\eqref{eq:Deltab2} becomes
\begin{equation}\label{eq:b2Sums}
\Delta b_2 = \frac{Q_\text{cm}}{Q_1} \sum_{E_\text{rel}} \left( e^{-E_\text{rel}\tilde{\omega}} - e^{-E_\text{rel}^{(0)}\tilde{\omega}} \right).
\end{equation}

In the case of a rotating gas both $Q_1$ and $Q_\text{cm}$ are determined from the single particle energy spectrum \eqref{eq:SingleParticleEnergies}. Including a factor of 2 to account for the spin states
\begin{equation} \label{eq:Q1}
Q_1 = \frac{2 e^{\tilde{\omega} \left(\eta + \xi+\frac{1}{2}\right)}}{\left(e^{\tilde{\omega}}-1\right) \left(e^{\eta \tilde{\omega}}-e^{\xi \tilde{\omega}}\right) \left(e^{\tilde{\omega} \left(\eta+\xi\right)}-1\right)}
\end{equation}
and $Q_\text{cm}=Q_1/2$. In the high temperature limit ($\tilde{\omega}\to 0$) the leading order behavior of Eq.~\eqref{eq:Q1} is exactly Eq.~\eqref{eq:NonIntQ1}. 

The magnetic field case is more complicated. As in the rotating case the same energy spectrum is used. However, $Q_1$ is obtained by exchanging $\eta$ and $\xi$ in Eq.~\eqref{eq:Q1} with $\eta_1$ and $\xi_1$. Similarly, $Q_\text{cm}$ is obtained by exchanging $\eta$ and $\xi$ with $\eta_\text{cm}$ and $\xi_\text{cm}$, and omitting the spin-counting factor of 2. 

To perform the remaining sums in Eq.~\eqref{eq:b2Sums} we need the eigenenergies of $H_\text{rel}$. Specifically, the case of $\eta=\eta_\text{rel}=1$, i.e.~the relative Hamiltonian is isotropic, is considered. The non-interacting ($a=0$) spectrum is $E_\text{rel}^{(0)}=2n+3/2$ and the spectrum in the unitary regime ($a\to\infty$) is $E_\text{rel}=2n+1/2$ \cite{Busch:1998}. For attractive interactions all states are included, but for repulsive interactions the $n=0$ bound state is omitted. 

For a rotating trapped gas in the high temperature limit with a large number of particles
\begin{alignat}{1}
\Delta b_2^\text{att} & = \frac{1}{4} - \frac{\tilde{\omega}^2}{32}  + \dots \\
\Delta b_2^\text{rep} & = -\frac{1}{4} - \frac{\tilde{\omega}}{4}  + \dots 
\end{alignat}
and for a trapped gas in a synthetic magnetic field
\begin{alignat}{1}
\Delta b_2^\text{att} & = \frac{\xi _1^2-\eta _1^2}{4 \left(\xi _{\text{cm}}^2-\eta _{\text{cm}}^2\right)} 
\nonumber \\ 
& + \frac{\left( \xi_1^2 - \eta_1^2 \right) \left(\eta_1^2 + \xi_1^2 - \eta_{\text{cm}}^2 - \xi_{\text{cm}}^2 - \tfrac{3}{2} \right) }{48 \left(\xi_{\text{cm}}^2 - \eta_{\text{cm}}^2\right)} \tilde{\omega}^2
+ \dots  \\
\Delta b_2^\text{rep} & = -\frac{\xi_1^2 - \eta_1^2}{4 \left(\xi_\text{cm}^2 - \eta_\text{cm}^2 \right)} + \frac{\left(\xi_1^2 - \eta_1^2 \right) \tilde{\omega}}{4 \left(\xi_\text{cm}^2-\eta_\text{cm}^2\right)} + \dots
\end{alignat}
We are now able to calculate thermodynamic quantities like the total energy and entropy of the gas from the grand potential Eq.~\eqref{eq:GTPOmega} \cite{HuangBook}.

To determine the energy and entropy the fugacity $z$ must be calculated first from $N(z)=-\pd\Omega/\pd\mu$, which is quadratic in $z$. In the rotating case there is always a single positive root of $N(z)$ and so the fugacity is always physical. In contrast, in the synthetic magnetic field case the roots are non-trivial. Despite the temperature {\it regime} relative to the trap ground-state energy being set by $\tilde{\omega}$, the temperature decreases as $\xi_1$ icreases, where  $\xi_1$ parameterizes the synthetic magnetic field. This is due to the restriction of isotropy in the relative coordinate, i.e.~$\eta_\text{rel}=1$ so that the transverse trapping must change as the magnetic field changes. For the attractive case, in all regimes there is a pole at $\xi_1=4/\sqrt{31} \approx 0.718$. For magnetic field strengths below this there is always one non-negative solution, which is physical. Above this value, we see that for larger $\tilde{\omega}$ more critical points appear at $\xi_1=4/\sqrt{28} \approx 0.756$ and $\xi_1=4/\sqrt{19} \approx 0.918$. Specifically, this occurs when $\tilde{\omega}>\left[(820+32\sqrt{651})/(9N) \right]^{1/3}$. Between these two critical points there is always one physical solution, but outside these values both solutions are complex. As a result of this behavior, we consider that the virial expansion for the synthetic magnetic field case in the attractive regime is valid for $\xi_1 \lesssim 0.75$. This problem is not present for the rotating gas where the parametrization merely restricts $\xi<1$, which is not an issue because we already expect that when $|\Omega_z|>\omega_z$ the gas becomes effectively untrapped. The same calculations can be performed for repulsive interactions to determine where the virial expansion is valid.

\begin{figure}[t!] 
\includegraphics[width=\columnwidth ]{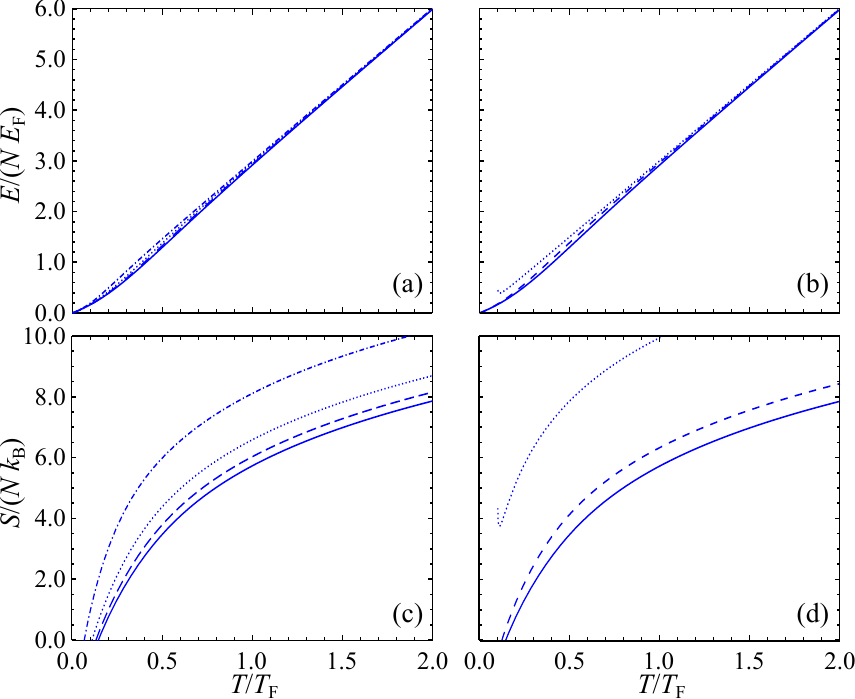}
\caption{Energy per particle (a), (b) and entropy per particle (c), (d) of a rotating atomic gas and a uniform synthetic magnetic field, respectively, for attractive interactions. Curves are plotted for $\xi=\xi_1=0.0 \, \, \text{(solid)}, 0.5 \, \, \text{(dashed)}, 0.75 \, \, \text{(dotted)}$ for both systems and additionally for $\xi=0.95 \, \, \text{(dot-dashed)}$ for the rotating system (a,c).}
\label{fig:EnergyAndEntropy}
\end{figure}

For the case of attractive interactions, we plot in Fig.~\ref{fig:EnergyAndEntropy} the energy (a,b) and entropy (c,d) per particle of a gas in a harmonic trap, subject to a rotation (a,c) or a synthetic magnetic field (b,d) in the case $\eta=\eta_\text{rel}=1$. 
Increasing the rotation or the synthetic magnetic field at a given temperature does not significantly affect the total energy of the system. In both cases the rotation and synthetic magnetic field can be viewed as a reparameterization of the transverse trapping frequencies, but at unitarity the interactions still dominate the energy of the system. In contrast, for both systems the entropy increases for larger $\xi$ (or $\xi_1$), as more states become accessible to the centre-of-mass because the rotation or synthetic magnetic field couple higher angular momentum states to lower energies. In the case of repulsive interactions, although the details are different, the same calculations can be performed and the results are qualitatively the same; the total energy of the gas is universal, but the entropy increases as rotation frequency or synthetic magnetic field strength is increased.

In conclusion, we have parameterized the problem of two ultracold atoms in a harmonic trap subject to a rotation or a synthetic magnetic field to retain spherical symmetry in the relative Hamiltonian. When the spherical symmetry is broken, even by a small perturbation, the $s$-wave contact interaction allows many more distinct relative energy states. These energy states appear because the spherical nature of the regularized interactions is incompatible with the cylindrical symmetry of the trap.  

In the special case of the relative Hamiltonian being isotropic in the unitary limit the total energy in the thermodynamic limit is universal and the entropy is not. This is due to the rotation or synthetic magnetic field coupling higher angular momentum states to lower energy levels and increasing the number of states available to the centre-of-mass of the system in a given energy range. However, in the unitary limit the total energy is dominated by the relative energy, which is determined by the interaction energy, and hence is largely independent of the rotation or synthetic magnetic field.



\bibliography{RotationAndMagField} 

\begin{thebibliography}{21}%
\makeatletter
\providecommand \@ifxundefined [1]{%
 \@ifx{#1\undefined}
}%
\providecommand \@ifnum [1]{%
 \ifnum #1\expandafter \@firstoftwo
 \else \expandafter \@secondoftwo
 \fi
}%
\providecommand \@ifx [1]{%
 \ifx #1\expandafter \@firstoftwo
 \else \expandafter \@secondoftwo
 \fi
}%
\providecommand \natexlab [1]{#1}%
\providecommand \enquote  [1]{``#1''}%
\providecommand \bibnamefont  [1]{#1}%
\providecommand \bibfnamefont [1]{#1}%
\providecommand \citenamefont [1]{#1}%
\providecommand \href@noop [0]{\@secondoftwo}%
\providecommand \href [0]{\begingroup \@sanitize@url \@href}%
\providecommand \@href[1]{\@@startlink{#1}\@@href}%
\providecommand \@@href[1]{\endgroup#1\@@endlink}%
\providecommand \@sanitize@url [0]{\catcode `\\12\catcode `\$12\catcode
  `\&12\catcode `\#12\catcode `\^12\catcode `\_12\catcode `\%12\relax}%
\providecommand \@@startlink[1]{}%
\providecommand \@@endlink[0]{}%
\providecommand \url  [0]{\begingroup\@sanitize@url \@url }%
\providecommand \@url [1]{\endgroup\@href {#1}{\urlprefix }}%
\providecommand \urlprefix  [0]{URL }%
\providecommand \Eprint [0]{\href }%
\providecommand \doibase [0]{http://dx.doi.org/}%
\providecommand \selectlanguage [0]{\@gobble}%
\providecommand \bibinfo  [0]{\@secondoftwo}%
\providecommand \bibfield  [0]{\@secondoftwo}%
\providecommand \translation [1]{[#1]}%
\providecommand \BibitemOpen [0]{}%
\providecommand \bibitemStop [0]{}%
\providecommand \bibitemNoStop [0]{.\EOS\space}%
\providecommand \EOS [0]{\spacefactor3000\relax}%
\providecommand \BibitemShut  [1]{\csname bibitem#1\endcsname}%
\let\auto@bib@innerbib\@empty
\bibitem [{\citenamefont {Chin}\ \emph {et~al.}(2010)\citenamefont {Chin},
  \citenamefont {Grimm}, \citenamefont {Julienne},\ and\ \citenamefont
  {Tiesinga}}]{Chin2010}%
  \BibitemOpen
  \bibfield  {author} {\bibinfo {author} {\bibfnamefont {C.}~\bibnamefont
  {Chin}}, \bibinfo {author} {\bibfnamefont {R.}~\bibnamefont {Grimm}},
  \bibinfo {author} {\bibfnamefont {P.}~\bibnamefont {Julienne}}, \ and\
  \bibinfo {author} {\bibfnamefont {E.}~\bibnamefont {Tiesinga}},\ }\href
  {http://link.aps.org/doi/10.1103/RevModPhys.82.1225} {\bibfield  {journal}
  {\bibinfo  {journal} {Rev. Mod. Phys.}\ }\textbf {\bibinfo {volume} {82}},\
  \bibinfo {pages} {1225} (\bibinfo {year} {2010})}\BibitemShut {NoStop}%
\bibitem [{\citenamefont {Shin}\ \emph {et~al.}(2008)\citenamefont {Shin},
  \citenamefont {Schunck}, \citenamefont {Schirotzek},\ and\ \citenamefont
  {Ketterle}}]{Shin:2008}%
  \BibitemOpen
  \bibfield  {author} {\bibinfo {author} {\bibfnamefont {Y.-i.}\ \bibnamefont
  {Shin}}, \bibinfo {author} {\bibfnamefont {C.~H.}\ \bibnamefont {Schunck}},
  \bibinfo {author} {\bibfnamefont {A.}~\bibnamefont {Schirotzek}}, \ and\
  \bibinfo {author} {\bibfnamefont {W.}~\bibnamefont {Ketterle}},\ }\href
  {http://dx.doi.org/10.1038/nature06473} {\bibfield  {journal} {\bibinfo
  {journal} {Nature}\ }\textbf {\bibinfo {volume} {451}},\ \bibinfo {pages}
  {689} (\bibinfo {year} {2008})}\BibitemShut {NoStop}%
\bibitem [{\citenamefont {Zwierlein}\ \emph {et~al.}(2006)\citenamefont
  {Zwierlein}, \citenamefont {Schunck}, \citenamefont {Schirotzek},\ and\
  \citenamefont {Ketterle}}]{zwierlein-2006-442}%
  \BibitemOpen
  \bibfield  {author} {\bibinfo {author} {\bibfnamefont {M.~W.}\ \bibnamefont
  {Zwierlein}}, \bibinfo {author} {\bibfnamefont {C.~H.}\ \bibnamefont
  {Schunck}}, \bibinfo {author} {\bibfnamefont {A.}~\bibnamefont {Schirotzek}},
  \ and\ \bibinfo {author} {\bibfnamefont {W.}~\bibnamefont {Ketterle}},\
  }\href {doi:10.1038/nature04936} {\bibfield  {journal} {\bibinfo  {journal}
  {Nature}\ }\textbf {\bibinfo {volume} {442}},\ \bibinfo {pages} {54}
  (\bibinfo {year} {2006})}\BibitemShut {NoStop}%
\bibitem [{\citenamefont {Busch}\ \emph {et~al.}(1998)\citenamefont {Busch},
  \citenamefont {Englert}, \citenamefont {Rzaewski},\ and\ \citenamefont
  {Wilkens}}]{Busch:1998}%
  \BibitemOpen
  \bibfield  {author} {\bibinfo {author} {\bibfnamefont {T.}~\bibnamefont
  {Busch}}, \bibinfo {author} {\bibfnamefont {B.-G.}\ \bibnamefont {Englert}},
  \bibinfo {author} {\bibfnamefont {K.}~\bibnamefont {Rzaewski}}, \ and\
  \bibinfo {author} {\bibfnamefont {M.}~\bibnamefont {Wilkens}},\ }\href@noop
  {} {\bibfield  {journal} {\bibinfo  {journal} {Foundations of Physics}\
  }\textbf {\bibinfo {volume} {28}},\ \bibinfo {pages} {549} (\bibinfo {year}
  {1998})}\BibitemShut {NoStop}%
\bibitem [{\citenamefont {Werner}\ and\ \citenamefont
  {Castin}(2006)}]{Werner2006}%
  \BibitemOpen
  \bibfield  {author} {\bibinfo {author} {\bibfnamefont {F.}~\bibnamefont
  {Werner}}\ and\ \bibinfo {author} {\bibfnamefont {Y.}~\bibnamefont
  {Castin}},\ }\href {http://link.aps.org/doi/10.1103/PhysRevA.74.053604}
  {\bibfield  {journal} {\bibinfo  {journal} {Phys. Rev. A}\ }\textbf {\bibinfo
  {volume} {74}},\ \bibinfo {pages} {053604} (\bibinfo {year}
  {2006})}\BibitemShut {NoStop}%
\bibitem [{\citenamefont {Kestner}\ and\ \citenamefont
  {Duan}(2007)}]{Kestner2007}%
  \BibitemOpen
  \bibfield  {author} {\bibinfo {author} {\bibfnamefont {J.~P.}\ \bibnamefont
  {Kestner}}\ and\ \bibinfo {author} {\bibfnamefont {L.-M.}\ \bibnamefont
  {Duan}},\ }\href {http://link.aps.org/doi/10.1103/PhysRevA.76.033611}
  {\bibfield  {journal} {\bibinfo  {journal} {Phys. Rev. A}\ }\textbf {\bibinfo
  {volume} {76}},\ \bibinfo {pages} {033611} (\bibinfo {year}
  {2007})}\BibitemShut {NoStop}%
\bibitem [{\citenamefont {Daily}\ and\ \citenamefont
  {Blume}(2010)}]{Daily2010}%
  \BibitemOpen
  \bibfield  {author} {\bibinfo {author} {\bibfnamefont {K.~M.}\ \bibnamefont
  {Daily}}\ and\ \bibinfo {author} {\bibfnamefont {D.}~\bibnamefont {Blume}},\
  }\href {http://link.aps.org/doi/10.1103/PhysRevA.81.053615} {\bibfield
  {journal} {\bibinfo  {journal} {Phys. Rev. A}\ }\textbf {\bibinfo {volume}
  {81}},\ \bibinfo {pages} {053615} (\bibinfo {year} {2010})}\BibitemShut
  {NoStop}%
\bibitem [{\citenamefont {Stöferle}\ \emph {et~al.}(2006)\citenamefont
  {Stöferle}, \citenamefont {Moritz}, \citenamefont {Günter}, \citenamefont
  {Köhl},\ and\ \citenamefont {Esslinger}}]{Staferle2006}%
  \BibitemOpen
  \bibfield  {author} {\bibinfo {author} {\bibfnamefont {T.}~\bibnamefont
  {Stöferle}}, \bibinfo {author} {\bibfnamefont {H.}~\bibnamefont {Moritz}},
  \bibinfo {author} {\bibfnamefont {K.}~\bibnamefont {Günter}}, \bibinfo
  {author} {\bibfnamefont {M.}~\bibnamefont {Köhl}}, \ and\ \bibinfo {author}
  {\bibfnamefont {T.}~\bibnamefont {Esslinger}},\ }\href
  {http://link.aps.org/doi/10.1103/PhysRevLett.96.030401} {\bibfield  {journal}
  {\bibinfo  {journal} {Phys. Rev. Lett.}\ }\textbf {\bibinfo {volume} {96}},\
  \bibinfo {pages} {030401} (\bibinfo {year} {2006})}\BibitemShut {NoStop}%
\bibitem [{\citenamefont {Liu}\ \emph {et~al.}(2009)\citenamefont {Liu},
  \citenamefont {Hu},\ and\ \citenamefont {Drummond}}]{Liu2009}%
  \BibitemOpen
  \bibfield  {author} {\bibinfo {author} {\bibfnamefont {X.-J.}\ \bibnamefont
  {Liu}}, \bibinfo {author} {\bibfnamefont {H.}~\bibnamefont {Hu}}, \ and\
  \bibinfo {author} {\bibfnamefont {P.~D.}\ \bibnamefont {Drummond}},\ }\href
  {http://link.aps.org/doi/10.1103/PhysRevLett.102.160401} {\bibfield
  {journal} {\bibinfo  {journal} {Phys. Rev. Lett.}\ }\textbf {\bibinfo
  {volume} {102}},\ \bibinfo {pages} {160401} (\bibinfo {year}
  {2009})}\BibitemShut {NoStop}%
\bibitem [{\citenamefont {Liu}\ \emph {et~al.}(2010)\citenamefont {Liu},
  \citenamefont {Hu},\ and\ \citenamefont {Drummond}}]{Liu2010}%
  \BibitemOpen
  \bibfield  {author} {\bibinfo {author} {\bibfnamefont {X.-J.}\ \bibnamefont
  {Liu}}, \bibinfo {author} {\bibfnamefont {H.}~\bibnamefont {Hu}}, \ and\
  \bibinfo {author} {\bibfnamefont {P.~D.}\ \bibnamefont {Drummond}},\
  }\href@noop {} {\bibfield  {journal} {\bibinfo  {journal} {Phys. Rev. A}\
  }\textbf {\bibinfo {volume} {82}},\ \bibinfo {pages} {023619} (\bibinfo
  {year} {2010})}\BibitemShut {NoStop}%
\bibitem [{\citenamefont {Daily}\ and\ \citenamefont
  {Blume}(2012)}]{Daily2012}%
  \BibitemOpen
  \bibfield  {author} {\bibinfo {author} {\bibfnamefont {K.~M.}\ \bibnamefont
  {Daily}}\ and\ \bibinfo {author} {\bibfnamefont {D.}~\bibnamefont {Blume}},\
  }\href {http://link.aps.org/doi/10.1103/PhysRevA.85.013609} {\bibfield
  {journal} {\bibinfo  {journal} {Phys. Rev. A}\ }\textbf {\bibinfo {volume}
  {85}},\ \bibinfo {pages} {013609} (\bibinfo {year} {2012})}\BibitemShut
  {NoStop}%
\bibitem [{\citenamefont {Nascimbène}\ \emph {et~al.}(2010)\citenamefont
  {Nascimbène}, \citenamefont {Navon}, \citenamefont {Jiang}, \citenamefont
  {Chevy},\ and\ \citenamefont {Salomon}}]{Nascimbene2010}%
  \BibitemOpen
  \bibfield  {author} {\bibinfo {author} {\bibfnamefont {S.}~\bibnamefont
  {Nascimbène}}, \bibinfo {author} {\bibfnamefont {N.}~\bibnamefont {Navon}},
  \bibinfo {author} {\bibfnamefont {K.~J.}\ \bibnamefont {Jiang}}, \bibinfo
  {author} {\bibfnamefont {F.}~\bibnamefont {Chevy}}, \ and\ \bibinfo {author}
  {\bibfnamefont {C.}~\bibnamefont {Salomon}},\ }\href
  {http://dx.doi.org/10.1038/nature08814} {\bibfield  {journal} {\bibinfo
  {journal} {Nature}\ }\textbf {\bibinfo {volume} {463}},\ \bibinfo {pages}
  {1057} (\bibinfo {year} {2010})}\BibitemShut {NoStop}%
\bibitem [{\citenamefont {Huang}\ and\ \citenamefont {Yang}(1957)}]{Huang1957}%
  \BibitemOpen
  \bibfield  {author} {\bibinfo {author} {\bibfnamefont {K.}~\bibnamefont
  {Huang}}\ and\ \bibinfo {author} {\bibfnamefont {C.~N.}\ \bibnamefont
  {Yang}},\ }\href {http://link.aps.org/doi/10.1103/PhysRev.105.767} {\bibfield
   {journal} {\bibinfo  {journal} {Phys. Rev.}\ }\textbf {\bibinfo {volume}
  {105}},\ \bibinfo {pages} {767} (\bibinfo {year} {1957})}\BibitemShut
  {NoStop}%
\bibitem [{\citenamefont {Idziaszek}\ and\ \citenamefont
  {Calarco}(2006{\natexlab{a}})}]{Idziaszek2006}%
  \BibitemOpen
  \bibfield  {author} {\bibinfo {author} {\bibfnamefont {Z.}~\bibnamefont
  {Idziaszek}}\ and\ \bibinfo {author} {\bibfnamefont {T.}~\bibnamefont
  {Calarco}},\ }\href {http://link.aps.org/doi/10.1103/PhysRevA.74.022712}
  {\bibfield  {journal} {\bibinfo  {journal} {Phys. Rev. A}\ }\textbf {\bibinfo
  {volume} {74}},\ \bibinfo {pages} {022712} (\bibinfo {year}
  {2006}{\natexlab{a}})}\BibitemShut {NoStop}%
\bibitem [{\citenamefont {Peng}\ \emph {et~al.}(2011)\citenamefont {Peng},
  \citenamefont {Li}, \citenamefont {Drummond},\ and\ \citenamefont
  {Liu}}]{Peng2011}%
  \BibitemOpen
  \bibfield  {author} {\bibinfo {author} {\bibfnamefont {S.-G.}\ \bibnamefont
  {Peng}}, \bibinfo {author} {\bibfnamefont {S.-Q.}\ \bibnamefont {Li}},
  \bibinfo {author} {\bibfnamefont {P.~D.}\ \bibnamefont {Drummond}}, \ and\
  \bibinfo {author} {\bibfnamefont {X.-J.}\ \bibnamefont {Liu}},\ }\href
  {http://link.aps.org/doi/10.1103/PhysRevA.83.063618} {\bibfield  {journal}
  {\bibinfo  {journal} {Phys. Rev. A}\ }\textbf {\bibinfo {volume} {83}},\
  \bibinfo {pages} {063618} (\bibinfo {year} {2011})}\BibitemShut {NoStop}%
\bibitem [{\citenamefont {W\'odkiewicz}(1991)}]{Wodkiewicz1991}%
  \BibitemOpen
  \bibfield  {author} {\bibinfo {author} {\bibfnamefont {K.}~\bibnamefont
  {W\'odkiewicz}},\ }\href {http://link.aps.org/doi/10.1103/PhysRevA.43.68}
  {\bibfield  {journal} {\bibinfo  {journal} {Phys. Rev. A}\ }\textbf {\bibinfo
  {volume} {43}},\ \bibinfo {pages} {68} (\bibinfo {year} {1991})}\BibitemShut
  {NoStop}%
\bibitem [{\citenamefont {Derevianko}(2003)}]{Derevianko2003}%
  \BibitemOpen
  \bibfield  {author} {\bibinfo {author} {\bibfnamefont {A.}~\bibnamefont
  {Derevianko}},\ }\href {http://link.aps.org/doi/10.1103/PhysRevA.67.033607}
  {\bibfield  {journal} {\bibinfo  {journal} {Phys. Rev. A}\ }\textbf {\bibinfo
  {volume} {67}},\ \bibinfo {pages} {033607} (\bibinfo {year}
  {2003})}\BibitemShut {NoStop}%
\bibitem [{\citenamefont {Pricoupenko}(2006)}]{Pricoupenko2006}%
  \BibitemOpen
  \bibfield  {author} {\bibinfo {author} {\bibfnamefont {L.}~\bibnamefont
  {Pricoupenko}},\ }\href {http://link.aps.org/doi/10.1103/PhysRevA.73.012701}
  {\bibfield  {journal} {\bibinfo  {journal} {Phys. Rev. A}\ }\textbf {\bibinfo
  {volume} {73}},\ \bibinfo {pages} {012701} (\bibinfo {year}
  {2006})}\BibitemShut {NoStop}%
\bibitem [{\citenamefont {Bolda}\ \emph {et~al.}(2003)\citenamefont {Bolda},
  \citenamefont {Tiesinga},\ and\ \citenamefont {Julienne}}]{Bolda2003}%
  \BibitemOpen
  \bibfield  {author} {\bibinfo {author} {\bibfnamefont {E.~L.}\ \bibnamefont
  {Bolda}}, \bibinfo {author} {\bibfnamefont {E.}~\bibnamefont {Tiesinga}}, \
  and\ \bibinfo {author} {\bibfnamefont {P.~S.}\ \bibnamefont {Julienne}},\
  }\href {http://link.aps.org/doi/10.1103/PhysRevA.68.032702} {\bibfield
  {journal} {\bibinfo  {journal} {Phys. Rev. A}\ }\textbf {\bibinfo {volume}
  {68}},\ \bibinfo {pages} {032702} (\bibinfo {year} {2003})}\BibitemShut
  {NoStop}%
\bibitem [{\citenamefont {Idziaszek}\ and\ \citenamefont
  {Calarco}(2006{\natexlab{b}})}]{Idziaszek2006a}%
  \BibitemOpen
  \bibfield  {author} {\bibinfo {author} {\bibfnamefont {Z.}~\bibnamefont
  {Idziaszek}}\ and\ \bibinfo {author} {\bibfnamefont {T.}~\bibnamefont
  {Calarco}},\ }\href {http://link.aps.org/doi/10.1103/PhysRevLett.96.013201}
  {\bibfield  {journal} {\bibinfo  {journal} {Phys. Rev. Lett.}\ }\textbf
  {\bibinfo {volume} {96}},\ \bibinfo {pages} {013201} (\bibinfo {year}
  {2006}{\natexlab{b}})}\BibitemShut {NoStop}%
\bibitem [{\citenamefont {Huang}(1963)}]{HuangBook}%
  \BibitemOpen
  \bibfield  {author} {\bibinfo {author} {\bibfnamefont {K.}~\bibnamefont
  {Huang}},\ }\href@noop {} {\emph {\bibinfo {title} {Statistical
  Mechanics}}},\ \bibinfo {edition} {2nd}\ ed.\ (\bibinfo  {publisher}
  {Wiley},\ \bibinfo {year} {1963})\BibitemShut {NoStop}%
\end{thebibliography}%

\end{document}